# The Simplest Solution to an Underdetermined System of Linear Equations

David Donoho

Department of Statistics

Stanford University

Hossein Kakavand, James Mammen

Department of Electrical Engineering

Stanford University

Email: {Donoho, hossein, jmammen}@stanford.edu

## Abstract

Consider a $d \times n$ matrix $A$, with $d < n$. The problem of solving for $x$ in $y = Ax$ is underdetermined, and has infinitely many solutions (if there are any). Given $y$, the minimum Kolmogorov complexity solution (MKCS) of the input $x$ is defined to be an input $z$ (out of many) with minimum Kolmogorov-complexity that satisfies $y = Az$. One expects that if the actual input is simple enough, then MKCS will recover the input exactly. This paper presents a preliminary study of the existence and value of the complexity level up to which such a complexity-based recovery is possible. It is shown that for the set of all $d \times n$ binary matrices (with entries 0 or 1 and $d < n$), MKCS exactly recovers the input for an overwhelming fraction of the matrices provided the Kolmogorov complexity of the input is $O(d)$. A weak converse that is loose by a $\log n$ factor is also established for this case. Finally, we investigate the difficulty of finding a matrix that has the property of recovering inputs with complexity of $O(d)$ using MKCS.

## I. INTRODUCTION

Many important technological problems require solutions to underdetermined systems of linear equations, i.e., systems of linear equations with fewer equations than unknowns. Examples arise in linear filtering, array signal processing, and inverse problems. For an underdetermined system of linear equations, if there is any solution, there are infinitely many solutions. In many applications, the "simplest" solution is most acceptable. Such a solution is inspired by the minimalist principle of Occam's Razor. For example, if the parameters of a model are being estimated then among all models that explain the data equally well, the one with the minimum number of parameters is most desirable. The model with the minimum number of parameters is, in fact, the sparsest solution. Thus in this example, one is looking for the sparsest solution and simplicity corresponds to sparseness.

An underdetermined system of linear equations, $y = Ax$, has infinitely many solutions (if there are any) and they form an affine space. The principle of Occam's Razor inspires choosing the simplest, i.e., the least complex solution in this affine space. In order to measure complexity, we use Kolmogorov complexity, which is a fundamental notion in algorithmic information theory. Clearly, if the Kolmogorov simplest solution is unique and $x$ is simple enough then it can be recovered from $y = Ax$ by choosing the simplest solution. The question then is: how simple should the input be so that it can be recovered as the simplest solution? Clearly if the input is sufficiently complex, it cannot be recovered using this approach of choosing the simplest solution. This suggests the existence of a complexity threshold below which all inputs can be recovered and above which recovery is not possible for all inputs. In this paper, we conduct a preliminary study and discover the existence of a complexity threshold for the family of $d \times n$ binary matrices.

The rest of the paper is organized as follows. In Section II, we present some preliminaries on Kolmogorov complexity and make precise the notion of complexity based recovery and complexity threshold. In Section III, we study the family of $d \times n$ binary matrices. We first show that any input sequence with complexity of $O(d)$ can be recovered for all but a vanishing fraction of these binary matrices. Thus most binary matrices allow complexity based recovery for inputs up to complexity of $O(d)$. We also derive a weak converse to this result that is off by a logarithmic factor. This establishes the complexity threshold for the family of $d \times n$ binary matrices to be roughly of order $d$. In Section IV, we explore the possibility of providing a concrete example of a matrix that allows complexity based recovery for inputs with complexity of $O(d)$ (since we know that many such matrices exist). Our results suggest that if a matrix can be easily described then complexity based recovery is not possible up to input complexity of $O(d)$. We specifically show such an example by constructing a family of binary matrices using Walsh functions.

## II. PRELIMINARIES

We first introduce the basic notions in algorithmic information theory used throughout the paper, including Turing machines, universal Turing machines, Kolmogorov complexity, and conditional Kolmogorov complexity. For formal definitions and comprehensive descriptions see [1], [4], [5]. Using these basic notions, we then define the minimum Kolmogorov complexity estimator in our setting, which is similar in spirit to the one in [2].

A *Turing machine* is a finite computing device equipped with an unbounded memory. There are several equivalent definitions and we use the following description suited to our purpose. The memory of the Turing Machine is given in the form of two tapes – an input tape and an output tape. Each tape is divided into an infinite number of cells in both directions. On each tape, a symbol from a finite alphabet $\Sigma$ can be written. $\Sigma$ contains a special symbol $B$ that denotes an *empty cell*. Let $\Sigma_0 = \Sigma \setminus B$. To access the information on the tapes, each tape is supplied by a read-write head. Time is discrete and at every step, each head sits on a cell of its corresponding tape. The heads are connected to a control unit with a finite number of states including a distinguished starting state "START" and a halting state "STOP".

In the beginning, the output tape is empty, i.e., each cell contains $B$. Whereas on the input tape a finite number of contiguous cells contain elements in $\Sigma_0$ and the rest are $B$. This finite sequence is called the *program* or the *input*. Initially, the control unit is in the "START" state, and the heads sit on the starting cells of the tapes. In every step, each head reads the symbol from the cell below it and sends it to the control unit. Depending on the symbols on the tapes and on its own state, the control unit carries out three basic operations:

- Send a symbol to each head to overwrite the symbol in the cell below it (the cell may also remain unchanged).
- Move each head one cell to the right or left or leave it unchanged.
- Make a transition into a new state (including the current state).

The machine halts when the control unit reaches the "STOP" state. The binary sequence present on the output tape when the Turing machine halts is called the *output* generated by that program. Alternatively one can say that the program on the input tape makes the Turing machine print the output. We further restrict the inputs to a Turing machine to be prefix-free so that no input that leads to a halting computation can be the prefix of any other such input.

A *Universal Turing Machine*, $U$ is a Turing machine that can simulate the behavior of any other Turing machine. It can be shown that such machines exist and can be constructed effectively (see Chapter 1 in [4]). For a comprehensive description of Turing machines and their properties see [4], [5].

The *Kolmogorov complexity* of a sequence $x$ chosen from the finite alphabet $\Sigma_0$ with respect to a universal Turing machine, $U$ is the length of the shortest program (input) that makes $U$ print $x$. The Invariance theorem states that the difference in the complexity of any sequence $x$ with respect to different universal Turing machines is at most a constant which is independent of $x$, see Section 7.2 in [5] for details. Hence, throughout the paper we shall fix the universal Turing machine $U$.

Let $s$ be a finite sequence of letters in $\Sigma_0$. The *conditional Kolmogorov complexity* of $x$ given $s$ with respect to $U$ is defined to be the length of the shortest program that makes $U$ print $x$ given that $U$ is provided with (knows) $s$. The invariance theorem stated above also holds for conditional complexity.

Now we explain what is meant by the Kolmogorov complexity of an $n$-tuple of real numbers. For an $n$-tuple $x = (x_1, \ldots, x_n) \in \mathbb{R}^n$, each coordinate $x_i, 1 \leq i \leq n$, consists of an integer part and a fractional part.

Suppose that the integer and fractional parts of all coordinates of $x$ have terminating binary expansions. Consider a Universal Turing machine with input alphabet $\{0, 1, B\}$ and output alphabet $\{0, 1, C, B\}$, where $C$ is just a symbol to denote a comma as a separator. A program is said to print $x$ if the Universal Turing machine halts for this program and the output tape contains the $2n$ binary strings denoting the integer and fractional parts of $x_1, \ldots, x_n$, in that order, separated by the symbol $C$. The conditional Kolmogorov complexity of $x$ given $n$ is the length of the shortest program that prints out $x$ and halts when the Universal Turing machine is provided with $n$. If any coordinate of $x$ does not have a terminating binary expansion then its Kolmogorov complexity is considered to be infinite. Throughout this paper, by $K(x)$ we shall mean the conditional Kolmogorov complexity of $x$ given $n$. Usually this is denoted as $K(x|n)$, however, since we do not deal with the unconditional version in this paper, for typographical ease we use $K(x)$.

A notion of estimation based on minimizing Kolmogorov complexity was introduced by Donoho in [2], in which, a signal corrupted by noise is estimated by simply choosing the signal with the least Kolmogorov complexity that fits the noise-corrupted signal to within the noise level. In this paper we shall extend this notion to recovering the input to an underdetermined system of linear equations given its output.

Consider the underdetermined system of equations, $y = Ax$, where $A \in \mathbb{R}^{d \times n}$ with $d < n$, $x \in \mathbb{R}^n$, and $y \in \mathbb{R}^d$. Our interest is in the problem of recovering the input $x$ from the output $y$ based on complexity.

We define the *Minimum Kolmogorov Complexity Solution (MKCS)* for the input given $y$ and $A$ to be

$$\hat{x}(y, A) = \arg_z \min\{K(z) : y = Az\}.$$

In words, $\hat{x}$ is the input $n$-tuple with the least Kolmogorov complexity among all inputs that would result in the same output $y$. For completeness we assume that if the minimizer is not unique then $\hat{x}$ is arbitrarily chosen, although this does not arise in our study.

An alternative way to look at MKCS would be as follows. Let $\mathcal{N}(A)$ denote the null-space of $A$ and let $y = Ax$. Then

$$\hat{x}(y, A) = \arg_z \min\{K(z) : z \in x + \mathcal{N}(A)\}.$$

Thus if $x$ is the actual input that resulted in output $y$, MKCS can recover it only if $x$ is the simplest element in $x + \mathcal{N}(A)$. Depending on the matrix $A$, $x$ may or may not be the simplest element in $x + \mathcal{N}(A)$. This raises the question of whether there exists a complexity threshold for recovery using MKCS for the given $A$. The answer to this question will depend on the specific matrix $A$ under consideration. Instead we study the complexity threshold for a family of matrices, which is the largest complexity level of the input for which it can be recovered using MKCS for almost all matrices in the family. This notion is made precise below. Let

$$K^*(A) = \max\{K : \hat{x} = x \text{ for all } x \text{ with } K(x) \leq K\}.$$

In words, given a matrix $A$, $K^*(A)$ is the largest complexity for which MKCS correctly recovers any input.

*Definition 1:* The *complexity threshold* for a family of $d \times n$ matrices, denoted as $\rho(d,n)$, is

$$\rho(d,n) = \max\{K : P\{K^*(A) \leq K\} \to 1\},$$

where the probability measure is the uniform measure on the assumed family of matrices.

In words, the complexity threshold of a family of $d \times n$ matrices is the highest complexity level such that for an overwhelming fraction of matrices in the given family, MKCS can correctly recover any input until that complexity level.

Note that, given the uniform probability measure, probabilistic statements about properties of $A$ are equivalent to statements about the fraction of matrices with that property in the family of $d \times n$ matrices under consideration.

As part of notation, we will use $c_i$ to denote constants that do not depend on $n$.

## III. RECOVERY FOR BINARY MATRICES

In this paper, we assume that we are given the output $y$, which arises due to the input $x$ and we want to recover the input based on complexity, i.e., using MKCS. In this section, we consider the family of $d \times n$ binary matrices. In Theorem 1, it is shown that for an overwhelming fraction of these matrices, the input can be recovered from the output using MKCS provided that the complexity of the input is $O(d)$. This means that $\rho(d,n) = \Omega(d)$. Theorem 2, following it, is a weak converse to Theorem 1 that is off by a $\log n$ factor. This establishes that $\rho(d,n) = O(d \log n)$.

*Theorem 1:* For an overwhelming fraction of $A \in \{0,1\}^{d \times n}$, $\hat{x} = x$ provided that $K(x) = O(d)$; more precisely, provided that $d - 2K(x) - c_1 = \omega(1)$, where $c_1$ is a positive constant independent of $x$ and $A$.

For the proof we need the following lemmas.

*Lemma 1:* If the null-space of a matrix $A \in \mathbb{R}^{d \times n}$ contains no non-zero element with complexity $\leq M$ then for any $x \in \mathbb{R}^n$ such that $K(x) \leq (M - c_2)/2$, $x$ is the only element in $x + \mathcal{N}(A)$ with complexity $\leq (M - c_2)/2$, where $c_2$ is a positive constant that does not depend on $x$ or $A$.

*Proof:* Observe that $K(x+y) \leq K(x) + K(y) + c_1$. This is because adding is a systematic operation which can be done using a program of constant length. As a result, one program to print $x + y$ would be to use the shortest programs for generating $x$ and $y$ and a wrapper code of some constant length $c_2$ to add them and then print the sum. Using the same idea for subtraction, it follows that $K(x - y) \leq K(x) + K(y) + c_2$. Hence we obtain, $K(y) \geq K(x - y) - K(x) - c_2$.

Now every $z \in x + \mathcal{N}(A)$, such that $z \neq x$, can be expressed as $x + y$ for some nonzero $y \in \mathcal{N}(A)$. Hence for all such $z$, $K(z) = K(x+y) \geq M - c_2 - K(x)$. This proves the lemma since $K(x) \leq (M - c_2)/2$. ∎

*Lemma 2:* For any non-zero $x \in \mathbb{R}^n$, the number of $b \in \{0,1\}^n$ such that $b \cdot x = \sum_{i=1}^n b_i x_i = 0$ is no more than $2^{n-1}$.

*Proof:* Define $B_x^n = \{b : b \cdot x = 0\}$. We shall prove that for any nonzero $x$, $|B_x^n| \leq 2^{n-1}$ by induction on $n$. It is easy to verify that the statement is true for $n = 2$.

Suppose for $n = k-1$, the statement is true. That is, $|B_x^{k-1}| \leq 2^{k-2}$ for all non-zero $x \in \mathbb{R}^{k-1}$. Now, for any non-zero $x \in \mathbb{R}^k$, we shall consider two cases based on $x_k$.

- Case 1: $x_k = 0$

  From our hypothesis for $n = k-1$, it follows that $|B_x^k| \leq 2.2^{k-2} = 2^{k-1}$.

- Case 2: $x_k \neq 0$

  This is further divided into two sub-cases.

  - Sub-case A: $b_k = 0$

    Denote the number of binary $k$-tuples such that $b.x = 0$ as $m$. Clearly $m \leq 2^{k-1}$.

  - Sub-case B: $b_k = 1$

    In this case $b \cdot x = 0$ implies that $\sum_{i=1}^{k-1} b_i x_i = -x_k \neq 0$. But $\sum_{i=1}^{k-1} b_i x_i = 0$ for $m$ of the $2^{k-1}$ binary $(k-1)$-tuples. Hence the number of binary $k$-tuples such that $b.x = 0$ is $\leq 2^{k-1} - m$.

Combining the above sub-cases, we obtain $|B_x^k| \leq m + 2^{k-1} - m$ for $x_k \neq 0$.

This completes the proof by induction for all $n$.

∎

Using the above two lemmas, we now prove Theorem 1.

*Proof:* Consider the set, $M_x$, of binary matrices that contain $x$ in their null-spaces. Thus $M_x = \{A : A \in \{0,1\}^{d \times n}, x \in \mathcal{N}(A)\}$ for any non-zero $x \in \mathbb{R}^n$. From Lemma 2, there are at most $2^{n-1}$ possibilities for a row of any $A \in M_x$. Hence, as each $A$ has $d$ rows, $|M_x| \leq 2^{(n-1)d}$.

Also, we know (using the same argument as in Theorem 7.2.4 of [1]) that for any complexity level $f(n)$, there are at most $2^{f(n)+1}$, $n$-tuples in $\mathbb{R}^n$ with $K(x) \leq f(n)$. Therefore there are at most $2^{f(n)+1} 2^{(n-1)d}$ binary matrices with a non-zero element of complexity $\leq f(n)$ in their null-spaces. Thus the fraction of such matrices, out of the total $2^{dn}$ matrices, goes to zero provided that $d - (f(n) + 1) \to \infty$ as $n \to \infty$; in other words, provided that $d - (f(n) + 1) = \omega(1)$.

Hence an overwhelming fraction of the matrices do not contain a non-zero element in the null-space with complexity $\leq f(n)$ provided that $d - f(n) - 1 = \omega(1)$. Assume that $f(n)$ satisfies this condition. Then from Lemma 1, it follows that for an overwhelming fraction of $A \in \{0,1\}^{d \times n}$, $x$ is the least complex element in $x + \mathcal{N}(A)$ provided that $K(x) \leq (f(n) - c_2)/2$. Clearly for any such $x$, $\hat{x} = x$. Thus we have shown that for an overwhelming fraction of binary matrices, $\hat{x} = x$, provided that $d - 2K(x) - c_2 - 1 = \omega(1)$. This proves the theorem.

∎

*A. Weak Converse*

In Theorem 1, we found that any input with complexity of $O(d)$ can be recovered for most binary matrices. A natural question that follows is beyond what complexity level can we no longer rely on a complexity based recovery. Obviously one would like this complexity level to coincide with that of Theorem 1 so that the complexity threshold is exactly determined. The following theorem states that for inputs with complexity of $O(d \log n)$, MKCS cannot recover at least one input for a constant fraction of the binary matrices. This results in an upper bound, $\rho(d,n) = O(d \log n)$.

*Theorem 2:* Let $d < n/\log n$ and $f(n) > d \log n + \log(1/(1-\alpha))$. Then there exists an $x \in \mathbb{R}^n$ with $K(x) \leq f(n) + c_3$ such that $\hat{x} \neq x$ for at least a fraction $\alpha \in (0,1)$ of $A \in \{0,1\}^{d \times n}$, where $c_3$ is a positive constant independent of $x$ and $A$.

*Proof:* There are $M = 2^{dn}$ distinct binary matrices. Let us label them $A_1, \ldots, A_M$. Now order the $m = 2^n$ binary $n$-tuples in lexicographical order and let $\{x_1, \ldots, x_m\}$ be this ordered collection. Let $y_{ij} = A_i x_j$ then $y_{ij}$ is a $d$-tuple where each of its $d$ elements can take values in $\{0, \ldots, n\}$. As a result, defining $S = \{y_{ij} : 1 \leq i \leq M, 1 \leq j \leq m\}$, we have $|S| \leq (n+1)^d$.

Consider the table with $y_{ij}$ in row $i$ and column $j$ as shown in Table I.

|       | $x_1$    | $x_2$    | $\cdots$ | $x_m$    |
|-------|----------|----------|----------|----------|
| $A_1$ | $y_{11}$ | $y_{12}$ | $\cdots$ | $y_{1m}$ |
| $A_2$ | $y_{21}$ | $y_{22}$ | $\cdots$ | $y_{2m}$ |
| $\vdots$ | $\vdots$ | $\vdots$ | $\vdots$ | $\vdots$ |
| $A_M$ | $y_{M1}$ | $y_{M2}$ | $\cdots$ | $y_{Mm}$ |

TABLE I

Consider row $i$ of this table. Suppose $y_{ij} = y_{ik}$ for some $1 \leq k < j$, i.e., entry $y_{ij}$ occurs in one of the prior columns of row $i$. Then by definition, MKCS cannot correctly recover $x_j$ for the matrix $A_i$. Next we show that for $f(n)$ big enough, there exists a column $j$ in the first $2^{f(n)}$ columns of the table ($j \leq 2^{f(n)}$) such that at least $\alpha$ fraction of its $M$ entries have occurred in some prior column. In what follows we will prove the contrapositive of this statement. That is, we will show that for the table restricted to $2^{f(n)}$ columns, if every column is such that, at most $\alpha$ fraction of its entries have occurred in prior columns then $2^{f(n)} < (n+1)^d/(1-\alpha)$.

Equivalently, we can ask the following question. If we impose the constraint that at most $\alpha$ fraction of entries in each column of a table are allowed to occur in prior columns, then using $|S| = (n+1)^d$ distinct elements, how many columns can such a table have? Consider a table with $N$ columns satisfying this constraint, then for each column at least $\bar{\alpha} = 1 - \alpha$ fraction of its entries have not occurred in prior columns. Place a mark on all such entries in each column. Since there are at least $\lceil \bar{\alpha} M \rceil$ entries in each column and $N$ columns, there are at least $\lceil \bar{\alpha} M \rceil N$ marks in the whole table. As a consequence, since there are $M$ rows, there exists a row that has at least $\lceil \bar{\alpha} M \rceil N / M \geq \bar{\alpha} N$ marks. But each mark in a row means that a new element of $S$, i.e., an element previously unused in the row, has been used. Since there are only $|S|$ elements, it must be that $\bar{\alpha} N \leq |S|$. That is, $N \leq |S|/\bar{\alpha}$. In our case this means that $2^{f(n)} \leq (n+1)^d/(1-\alpha)$. That is, $f(n) \leq d \log n + \log(1/(1-\alpha))$.

Thus we have shown that if $f(n) > d \log n + \log(1/(1-\alpha))$, there exists a binary $n$-tuple $x_i$ such that $\hat{x} \neq x$ for at least a fraction $\alpha$ of $A \in \{0,1\}^{d \times n}$, Moreover the complexity of $x_i$ for $i = 2^{f(n)}$ can be shown to be no more than $f(n) + c_3$.

Thus we have shown that if $f(n) > d \log n + \log(1/(1-\alpha))$, there exists a binary $x \in \mathbb{R}^n$ with $K(x) \leq f(n) + c_3$ such that for $\alpha$ fraction of the binary matrices, $\hat{x} \neq x$. ∎

The above theorem is a weak statement about the performance limit of MKCS due to the following two reasons. First, it differs from what can be achieved in Theorem 1 using MKCS by a factor of $\log n$. Second, and more importantly, it only says that at least for one input, MKCS does not provide the correct answer. A more satisfactory converse should provide the complexity level beyond which MKCS cannot recover at least a constant fraction of all inputs below that complexity level.

## IV. RECOVERY FOR MATRICES THAT CAN BE EASILY DESCRIBED

As shown in Theorem 1, the complexity threshold for large $d \times n$ binary matrices is $O(d)$. The proof of this theorem relies on the fact that an overwhelming fraction of the matrices do not have non-zero elements simpler than $O(d)$ in their null-spaces. It is natural to try and construct a matrix with this property. As a first attempt, consider the following $d \times n$ binary matrices that are easy to describe, i.e., matrices that have a "name": the matrix of all zeros or the matrix of all ones or the submatrix consisting of the first $d$ rows of the identity matrix. It is easy to check that all these matrices have simple non-zero elements in their null-spaces, as a result of which none of these matrices can recover all inputs with $O(d)$ complexity. In what follows, we show that such a result holds for a large family of matrices. Specifically, we consider binary matrices whose rows are Walsh functions. The Walsh functions can be generated recursively and hence are easy to describe, which means that they have low Kolmogorov complexity. We show that if the complexity of the input exceeds $\log n$, then at least for a constant fraction of the matrices, MKCS cannot recover a constant fraction of the inputs correctly.

In order to define the family of matrices under consideration, we first define Walsh functions. Recall that a Hadamard matrix $H$ of order $n$ is an $n \times n$ matrix of $+1$s and $-1$s such that $HH^T = nI$, where $I$ is the identity matrix (see [6], Chapter 6.1). Thus the rows of a Hadamard matrix are orthogonal. The Sylvester construction provides a way to generate Hadamard matrices of order $n$ for all $n$ that are powers of 2. According to the

Sylvester construction, if $H_n$ is a Hadamard matrix then so is

$$H_{2n} = \begin{bmatrix} H_n & H_n \\ H_n & -H_n \end{bmatrix}.$$

The $2^n$ Walsh functions of order $n$ are given by the rows of the Hadamard matrix of order $2^n$ when arranged in so-called "sequency" order. The sequency of a sequence of $+1$s and $-1$s is the number of sign changes in the sequence. Since the rows of a Hadamard matrix are orthogonal, it follows that the Walsh functions are also orthogonal.

Let $w_i, 1 \leq i \leq n$, denote Walsh functions of order $n$. A program to generate $w_i$ would be to generate the Hadamard matrix of order $n$ using the recursive formula given by the Sylvester construction and arrange the rows in sequency order and then print out row $i$ of this matrix. The index $i \leq n$ needs to be specified in the program which consumes no more than $\log n$ bits. The remaining part of the program has some constant length $c_4$. Thus the Kolmogorov complexity of any Walsh function, $K(w_i) \leq \log n + c_4$.

Consider a $d \times n$ matrix whose rows are $d$ distinct Walsh functions. In our setting, the order of the rows does not matter and so we shall assume that the rows of $A$ are naturally ordered according to the ordering on $\{w_i\}$. That is, if row $i$ equals $w_{k(i)}$ and row $j$ equals $w_{k(j)}$ then $i < j$ implies that $k(i) < k(j)$. Thus there are $M = \binom{n}{d}$ such distinct matrices. Let $\mathcal{A}$ denote the set of these $M$ matrices and consider the set

$$\mathcal{S}(K) = \{(A, x, y) : A \in \mathcal{A}, K(x) \leq K, y = Ax\}.$$

This is the set of triples (matrix, input, output) with input Kolmogorov complexity $\leq K$.

*Theorem 3:* For the set of triples $\mathcal{S}(K)$ defined above, if $K = \Omega(\log n)$ then there exists a constant $\eta > 0$, such that a fraction $\eta$ of the triples have $\hat{x}(A, y) \neq x$.

*Proof:* In what follows, we show that a (strictly positive) constant fraction of the triples $(A, x, y) \in \mathcal{S}(K)$ have $y = 0$. It is then easy to see that this proves the theorem since for each matrix, among all inputs that result in output $0$, only one can be recovered.

Recall that $K(w_i) \leq \log n + c_4$ for all Walsh functions $w_i$. Also there are at most $2^{K+1}$ $n$-tuples whose Kolmogorov complexity is no more than $K$. Given a number $c$ with complexity $K(c)$, consider the $n$-tuple $cw_i$, which is a multiple of a Walsh function. Since multiplication can be accomplished using a program of constant length $c_5$ (say), its Kolmogorov complexity, $K(cw_i) \leq \log n + c_4 + c_5 + K(c)$.

Now given complexity level $K$, there are at least $\alpha 2^K (\alpha > 0)$, numbers with Kolmogorov complexity not exceeding $K$. As a consequence, there are at least $\alpha n 2^K$ $n$-tuples that are both multiples of Walsh functions and have Kolmogorov complexity at most $\log n + c_4 + c_5 + K$. If $K = \Omega(\log n)$ then by rewriting this it follows that there are at least $\alpha 2^{K-c_4-c_5}$ $n$-tuples that are multiples of Walsh functions and have Kolmogorov complexity at most $K$. For each such $n$-tuple $x$, $\binom{n-1}{d}$ of the matrices in $\mathcal{A}$ result in $y = Ax = 0$ due to the orthogonality of the Walsh functions. But

$$\binom{n-1}{d} / \binom{n}{d} = 1 - \frac{d}{n},$$

and hence at least $(1 - \frac{d}{n})\alpha 2^{-c_4-c_5} > 0$ fraction of the $(A, x, y)$ triples result in $y = 0$. ∎

Above, we considered the triples $(A, x, y)$ but this is equivalent to considering the pairs $(A, x)$ since $y = Ax$.

We are interested in finding a matrix with the property that MKCS recovers most inputs with complexity of $O(d)$. The above theorem shows that at least a constant fraction of the matrices having Walsh functions as their rows do not have this property. Although unlikely, this theorem does leave the possibility that some of these matrices may have the desired property. This raises the question whether it is possible to specify a matrix that can recover all inputs with $O(d)$ complexity. At least no simple description of such a matrix seems likely. Although, as we showed earlier, most random matrices that are hard to describe do have this property.

## V. Discussion and Conclusion

The results obtained in our study have striking similarities to the results in [3] regarding sparse solutions of underdetermined systems of linear equations. In [3], it was shown that if the input is sufficiently sparse, i.e., has sufficiently few nonzero elements, the solution to the $\ell^1$ minimization problem is the unique sparsest solution. As a specific example, if $A$ is a uniform random orthoprojector from $\mathbb{R}^n$ to $\mathbb{R}^d$, such that $d/n$ is a constant then for almost all such $A$, given $y$, a sparse input $x$ with less than $d/2$ nonzero elements can be recovered exactly with $\ell^1$ minimization over all the solutions of $y = Ax$. See Corollary 1.5 in [3] for details.

The above result says that if $x$ is sparse enough then it can be recovered from $y = Ax$ by choosing the sparsest solution. Note the correspondence of the sparsity result and Theorem 1. Roughly speaking, the sparsity result shows that inputs with less than $d/2$ nonzero elements can be uniquely recovered based on sparsity for an overwhelming fraction of $d \times n$ orthoprojection matrices. Similarly the complexity result states that inputs with complexity of $O(d)$ can be uniquely recovered based on complexity for an overwhelming fraction of $d \times n$ binary matrices.

In this paper, we made a preliminary study of recovery in underdetermined systems based on Kolmogorov complexity. We showed the existence of a complexity threshold for the family of binary matrices, i.e. we showed that for most large binary $d \times n$ matrices ($d < n$), an input can be recovered based on complexity as long as its complexity is $O(d)$. We also showed that this complexity based recovery fails for at least one input with complexity of $O(d \log n)$ for a constant fraction of the binary matrices. The correspondence of results based on Kolmogorov complexity with those based on sparsity suggests that Kolmogorov complexity does capture the intuitive notion of simplicity that matches well with sparsity. We also explored the possibility of specifying a matrix that allows recovery of inputs with complexity of $O(d)$. Our initial results suggest that matrices that have a simple description may not possess this property.